\documentclass{elsart}
\usepackage{amssymb}
\usepackage{amsmath}
\usepackage{graphicx}
\newcommand{\bmath}[1]{\mbox{\boldmath ${#1}$}}
\newcommand{\dd}{\mbox{\rm d}}
\begin{document}
\begin{frontmatter}

\title{\large{\bf Model-independent analysis of the neutron-proton
final-state interaction region in the $\bmath{pp\to pn\,\pi^+}$ reaction}}

\author[Dubna,Almaty]
{Yuri N.\ Uzikov}~\thanks{E-mail:~uzikov@nusun.jinr.dubna.su}
and\ \ \
\author[UCL]{Colin Wilkin}~\thanks{Corresponding author:
C.~Wilkin, Physics and Astronomy Department,
UCL, Gower Street, London WC1E 6BT, UK.\ \ \ 
E-mail:~cw@hep.ucl.ac.uk}

\address[Dubna]{JINR, LNP, Dubna, 141980 Russia}
\address[Almaty]{Kazakh State University, Almaty, 480121 Kazakhstan}
\address[UCL]{University College London, London, WC1E 6BT, UK}

\begin{abstract}
Experimental data on the $pp\to pn\,\pi^+$ reaction measured in an
exclusive two-arm experiment at 800~MeV show a narrow peak arising
from the strong proton-neutron final-state interaction. It was claimed,
within the framework of a certain model, that this peak contained up
to a 25\% spin-singlet final state contribution. By comparing the data 
with those of $pp\to d\,\pi^+$ in a largely model-independent way, 
it is here demonstrated that at all the angles measured the whole of 
the peak could be explained as being due to spin-triplet final states, 
with the spin-singlet being at most a few percent. Good qualitative 
agreement with the measured proton analysing power is also found 
within this approach.
\end{abstract}

\end{frontmatter} 

%%%%%%%%%%%%%%%%%%%%%%%%%%%%%%%%%%%%%%%%%%%%%%%%%%%%%%%%%%%%%%%%%%%%%%
%%%%%%%%%%%%%%%%%%%%%%%%%%%%%%%%%%%%%%%%%%%%%%%%%%%%%%%%%%%%%%%%%%%%%%
\newpage

Inclusive data on the $pp\to \pi^+X$ reaction in the 400~MeV to
1~GeV range generally show two structures. There is firstly a broad peak
corresponding to the quasi-free production of the $\Delta$-isobar. In 
addition, there is an enhancement near the edge of phase space arising
from the strong neutron-proton final-state interaction (fsi) in either the 
spin-triplet or singlet $S$-wave which comes when the $np$ excitation
energy, $E_{np}$, is only a few MeV. The details of the enhancement 
region are hard to investigate in a single-arm experiment, where 
only the $\pi^+$ is measured, because of
contamination from the much larger two-body $pp\to d\,\pi^+$ 
reaction~\cite{Falk}. The simplest way to overcome this background
is by measuring in coincidence the final proton and pion in
the exclusive $pp\to pn\,\pi^+$ reaction. The most complete examples
of such an experiment were carried out at LAMPF, where both the
five-fold differential cross section~\cite{HG} and the proton
analysing power~\cite{Hancock} were studied at a proton beam
energy of $T_p=800$~MeV.

LAMPF data taken at one pair of proton/pion angles are shown in Fig.~1
as a function of the detected proton momentum~\cite{HG}. The large peak
on the right is associated with protons and pions formed in the 
$pp\to \Delta^{++}n/\Delta^{+}p$ reactions and its magnitude and width
can be explained in different versions of
one-meson-exchange models~\cite{HG,Yuri}.
At the maximum of the smaller peak, $E_{np}$ is about 1~MeV and the
cross section is strongly influenced by the $np$ fsi. The form of this
peak is normally parameterised by Watson final-state interaction
factors~\cite{Watson,GW} which take into account the nearby poles in the
scattering amplitudes due to the deuteron bound state in the
spin-triplet case and the anti-bound state for the spin-singlet.
Since the latter is closer to $E_{np}=0$, the singlet fsi peak
is expected to be narrower than that of the triplet. Though this difference
in shape can, in principle, be used to extract the relative amount of the
$np$ spin-triplet and singlet in the final-state peak, the limited number
of points in the peak and the modest resolution in $E_{np}$ makes this
impractical for the LAMPF data.

In analysing their experiment, the authors of Ref.~\cite{HG} made the
\textit{ad hoc} assumption that the one-meson-exchange model for
quasi-free $\Delta$ production was valid for $E_{np}$ above 10~MeV
and that, below this excitation energy, the prediction could be smoothly
joined onto Watson fsi factors. The relative spin-singlet strength depends, of
course, upon the kinematics and their analysis suggested that the 
fraction was about 15-25\% of the total for the different angle pairs 
measured. This is close to a statistical mixture of 25\% and
is in complete contrast to measurements at lower energies, where the
singlet fraction is typically 10\%~\cite{FW2}. Alternative
model-dependent analyses of the LAMPF data do lead to smaller singlet
production, but they depend upon other assumptions made~\cite{Dubach}.
In view of these differing conclusions, it is worthwhile to seek a
different way of deducing the spin-singlet contribution from these data.

When the square of the low energy $np$ triplet $S$-state scattering
wave function at energy $E_{np}= k^2/m_N$, where $m_N$ is twice the
$n$-$p$ reduced mass, is analytically continued to negative energy,
it manifests the deuteron pole at $k^2=-\alpha_t^2$, where
$\alpha_t=0.232$~fm$^{-1}$. It has recently been shown~\cite{FW1}
that the relative
normalisation of the scattering, $\Psi_k(r)$, and bound-state,
$\Psi_{\alpha_t}(r)$, wave function depends purely upon the deuteron
binding energy and is independent of the $np$ potential~\cite{FW1}.
Using real boundary conditions, it follows that at short distances
\begin{equation}
\label{FW}
\left[\Psi_k(r)\right]^2 \approx
\frac{2\pi}{\alpha_t(k^2+\alpha_t^2)}\,
\left[\Psi_{\alpha_t}(r)\right]^2\:.
\end{equation}
Apart from questions associated with the $D$-state contributions which
are small at low $k^2$, this relation becomes exact as $k^2\to -\alpha_t^2$.
In the scattering region where $E_{np} \geq 0$, it allow one to
estimate the spin-triplet amplitude for $pp\to pn\,\pi^+$
in terms of that for $pp\to d\,\pi^+$. Thus, for any value of the
initial ($\sigma_i$) and final spin-triplet projection $\lambda$, the
matrix elements are related by
\begin{equation}
\label{ME}
M_{\sigma_1\sigma_2}^{\lambda}(pp\to\{np\}_t\,\pi^+)
\approx f(k^2)\,M_{\sigma_1\sigma_2}^{\lambda}(pp\to d\,\pi^+)\:,
\end{equation}
where
\begin{equation}
\label{fsi}
f^2(k^2) = \frac{2\pi\,m_N}{\alpha_t(k^2+\alpha_t^2)}\:\cdot
\end{equation}
We are here using a normalisation where the unpolarised
$pp\to d\,\pi^+$ cross section is given by
\begin{equation}
\label{pid}
\frac{\dd\sigma}{\dd\Omega^*}= \frac{1}{64\pi^2s_{pp}}\,
\frac{q_{\pi d}^*}{q_{p}^*}\,\frac{1}{4}\,\sum_{\sigma_1\sigma_2\lambda}
\mid M_{\sigma_1\sigma_2}^{\lambda}(pp\to d\,\pi^+)\mid^2\:,
\end{equation}
$s_{pp}$ is the square of the centre-of-mass (cm) energy, and $q_p^*$ and
$q_{\pi d}^*$ are the cm momenta in the initial and final states
respectively. At low $E_{np}$ the approximation of eq.~(\ref{ME})
reproduces very well
the results of single-arm experiments where only the pion is
detected~\cite{FW2}.

We now extend this approach to treat two-arm experiments. Starting from
eq.~(\ref{ME}), the triplet contribution to the laboratory five-fold 
differential cross section for the detection of a pion at an angle 
$\theta_{\pi}$ and a proton at an angle $\theta_p$ becomes
\begin{equation}
\label{yuri}
\frac{\dd^5\sigma_t(pp\to pn\pi^+)}{\dd p_p\,\dd\Omega_p\,\dd\Omega_\pi}
= \frac{1}{16 \pi^3}\, s_{pp} \frac{q_{p}^*}{q_{\pi d}^*}\,\Phi \,f^2(k^2)\,
\frac{\dd\sigma}{\dd\Omega_{\pi}^*}(pp\to d\pi^+)\:,
\end{equation}
where the phase-space factor is
\begin{equation}
\label{PS}
\Phi=\frac{p_p^2 p_\pi^3}{ p_0\, m_p\, E_p\,
|p_\pi^2\, E_n -{\bf p}_\pi\cdot {\bf p}_n E_\pi|}\:\cdot
\end{equation}
Here $p_0$ is the beam momentum, $E_i$ and $p_i$ ($i=p,n,\pi$) are the
laboratory energy and momentum of the $i$-th particle in the final state.

The pion production angles in the laboratory ($\theta_\pi$) and  cm
($\theta_{\pi}^*$) systems are related by
\begin{equation}
\label{theta}
E_0\,E_\pi - p_0\,p_\pi \cos{\theta_\pi}=
\varepsilon_0\,\varepsilon_\pi-q_p^*\, q_\pi^*
\cos{\theta_\pi^*}\:,
\end{equation}
where $\varepsilon_\pi (\varepsilon_0)$  and $q_\pi^*$  are the energy 
of the pion (incident proton) and 3-momentum of the
pion in the overall cm system and $E_0$ is the total laboratory energy of the
incident  proton. Due to the difference between the effective mass of 
the final $pn$ state and that of the deuteron, 
$q_{\pi}^*\not =q_{\pi d}^*$. However, this effect is quite small at 
low $E_{np}$.

The results of eq.~(\ref{yuri}), which should be valid at low relative
energies $E_{np}$, do not depend upon the details of the pion production
dynamics and automatically include the final state interaction in the
triplet $pn$ system.

The input $pp\to d\,\pi^+$ cross sections are taken from the SAID
SP96 parameterisation~\cite{SAID} and this procedure should involve 
errors that are smaller than those of an individual experiment.
The predictions of our approach for
the triplet $pp\to pn\,\pi^+$ cross section are shown at one
angle pair in Fig.~1 together with an evaluation of quasi-free
$\Delta$-production~\cite{Yuri}. The values of the $np$ excitation energies
are indicated and from this it is seen that when $E_{np}$ is below
about 5~MeV the magnitude and shape of the fsi peak are both well
reproduced. For comparison the fsi parameterisation of Goldberger 
and Watson~\cite{GW}, multiplied by phase space and normalised to the 
peak value, is also illustrated.

Results at different angles in the fsi region are shown in Fig.~2.
Larger deviations from the predictions are to be found when the minimum
value of the excitation energy, $E_{np}^{min}$, is increased. This is the
case at $(\theta_p,\theta_{\pi}) = (25^{\circ},40^{\circ})$ and
$(30^{\circ},28^{\circ})$ where $E_{np}^{min}$ are respectively
13 and 28~MeV. Even in this last case, the changes induced by the
modified kinematics are quite small, as can be judged from the dot-dash
curve shown in the figure. A potentially more serious effect arises from
averaging the estimates over the experimental angular acceptance. As
shown in Fig.~2, this tends to reduce a little the predictions when 
$E_{np}^{min}\leq 1$~MeV but can increase them otherwise. After
smearing, there seems to be some underprediction in Fig.~2b, though it 
should be stressed that these data were taken from the polarisation
experiment~\cite{Hancock} where the consistency with the earlier 
unpolarised cross section run was found only on the 15\% level. 

Since the predictions of the F\"aldt-Wilkin extrapolation
theorem~\cite{FW1} reproduce most of the magnitude and angular dependence of
the cross section leading to the fsi peaks, this is strong evidence
that the vast bulk of the LAMPF data corresponds to $np$ triplet final
states. Extra confirmation of this interpretation is found from the
proton analysing power which was measured at two pairs of angles in
the fsi region~\cite{Hancock}. In the fsi peak we expect $A_y$ to be
essentially constant at the value corresponding to that of
$pp\to d\,\pi^+$ at the appropriate pion angle. These values are
shown with the experimental data in Fig.~3. Though at
$(14.5^{\circ},21^{\circ})$ the data seem to fall a little above
the prediction, deviations of this size are not unknown in polarisation
measurements.

Although we have reproduced well the LAMPF data at low $E_{np}$ with
just triplet terms,
it may be helpful to try to estimate an upper bound on the possible
spin-singlet contribution from the areas under the peaks. 
The overall systematic error due to beam
normalisation and detector efficiency in the measurement was estimated
to be about 7\%~\cite{HG}, to which must be compounded some error
coming from the $pp\to d\,\pi^+$ input~\cite{SAID}. The error arising 
from using the extrapolation theorem in the scattering domain is likely to be
rather smaller than this, with variations in the wave functions at short
distances being on the 1--2\% level for $E_{np}<10$~MeV~\cite{FW3}.
Further work is needed to include the $D$-state effects more consistently,
though it has been shown that the extrapolation theorem is valid in
this coupled-channel case~\cite{Smirnov}. Under the present conditions,
we would estimate the accuracy of eq.~(\ref{yuri}) to be better than 5\%
for $E_{np}\leq 3$~MeV~\cite{smuzikov}. We therefore conclude that the 
singlet contribution to the unpolarised LAMPF data at low $E_{np}$ 
in Fig.~1 is below about 10\%. This upper bound is about a factor of two
less than the average quoted in the experimental paper~\cite{HG},
though it must be stressed that this involved considerable model
dependence, including the choice of a 10 MeV matching point.

There is a 15\% normalisation uncertainty between the first
measurements of the unpolarised cross section at LAMPF~\cite{HG} and their
later experiment, where the primary purpose was the determination of the
analysing power~\cite{Hancock}. If, nevertheless, we take seriously
the disagreement with the latter data shown in fig.~2b, it is possible that
this is due to a spin-singlet contribution. Assuming that
the singlet cross section to be of the form
\begin{equation}
\label{singlet}
\frac{\dd^5\sigma_s(pp\to pn\pi^+)}{\dd p_p\,\dd\Omega_p\,\dd\Omega_\pi}
=\zeta\,\frac{(k^2+\alpha_t^{\,2})}{(k^2+\alpha_s^{\,2})}\times
\frac{\dd^5\sigma_t(pp\to pn\pi^+)}{\dd p_p\,\dd\Omega_p\,\dd\Omega_\pi}\:,
\end{equation}
where $\alpha_s=-0.04$~fm$^{-1}$ then, after averaging over
acceptance, a value of $\zeta=0.03$ restores the agreement with the
data. This would correspond to a singlet fraction integrated over the 
range $0\leq E_{np}\leq 3$~MeV of about 10\%. However, such a fraction
would make the agreement worse in fig.~2c, where the data were taken
at a rather similar cm angle but in the other hemisphere. 

To improve the sensitivity to the singlet/triplet ratio using the
extrapolation theorem, the ratio of the $pp\to pn\,\pi^+$ and
$pp\to d\,\pi^+$ cross sections has to be established better by
measuring both reactions in the same experiment. Data with better
resolution on $E_{np}$ would also allow one to investigate the
singlet/triplet ratio from the shape of the fsi peak. A new measurement
of the $pp\to pn\,\pi^+$ reaction at the ANKE spectrometer of the
proton synchrotron COSY-J\"ulich, where both protons
and pions were detected near the forward direction at $T_p=492$~MeV,
involved both these improvements and can therefore put more stringent
limits on the spin ratio~\cite{VK}.

In conclusion, we have generalised the use of the extrapolation theorem
linking $np$ scattering and bound-state wave functions to describe
five-fold differential cross sections. In so doing, we have
shown that the LAMPF $pp\to pn\,\pi^+$ cross section data at 800~MeV
are consistent with there being no final-state spin-singlet
contribution. This smallness should not come as a
complete surprise. If one assumes that pion 
production passes through an intermediate $\Delta$ 
isobar~\cite{HG,Yuri,Dubach}, then the $S$-wave $\Delta\,N$ 
intermediate state cannot lead to singlet $np$ final states. These
must be produced through higher partial waves or from non-resonant 
pion production.

This work was initiated within the framework of the ANKE collaboration
at COSY and the authors would like to thank all their colleagues, especially
V.~Abaev, V.I.~Komarov, V.~Koptev, and V.~Kurbatov, for many fruitful 
discussions. Financial support from the Forschungszentrum J\"ulich and
BMBF (WTZ grant KAZ 99/001) and the generous hospitality of
Institut f\"ur Kernphysik, where the work was carried out, are gratefully
acknowledged.
%%%%%%%%%%%%%%%%%%%%%%%%%%%%%%%%%%%%%%%%%%%%%%%%%%%%%%%%%%%%%%%%%%%%%%
%%%%%%%%%%%%%%%%%%%%%%%%%%%%%%%%%%%%%%%%%%%%%%%%%%%%%%%%%%%%%%%%%%%%%%

%%%%%%%%%%%%%%%%%%%%%%%%%%%%%%%%%%%%%%%%%%%%%%%%%%%%%%%%%%%%%%%%%%%%%%
%%%%%%%%%%%%%%%%%%%%%%%%%%%%%%%%%%%%%%%%%%%%%%%%%%%%%%%%%%%%%%%%%%%%%%
\input epsf
\begin{figure}[t]
\begin{center}
\mbox{\epsfxsize=4in \epsfbox{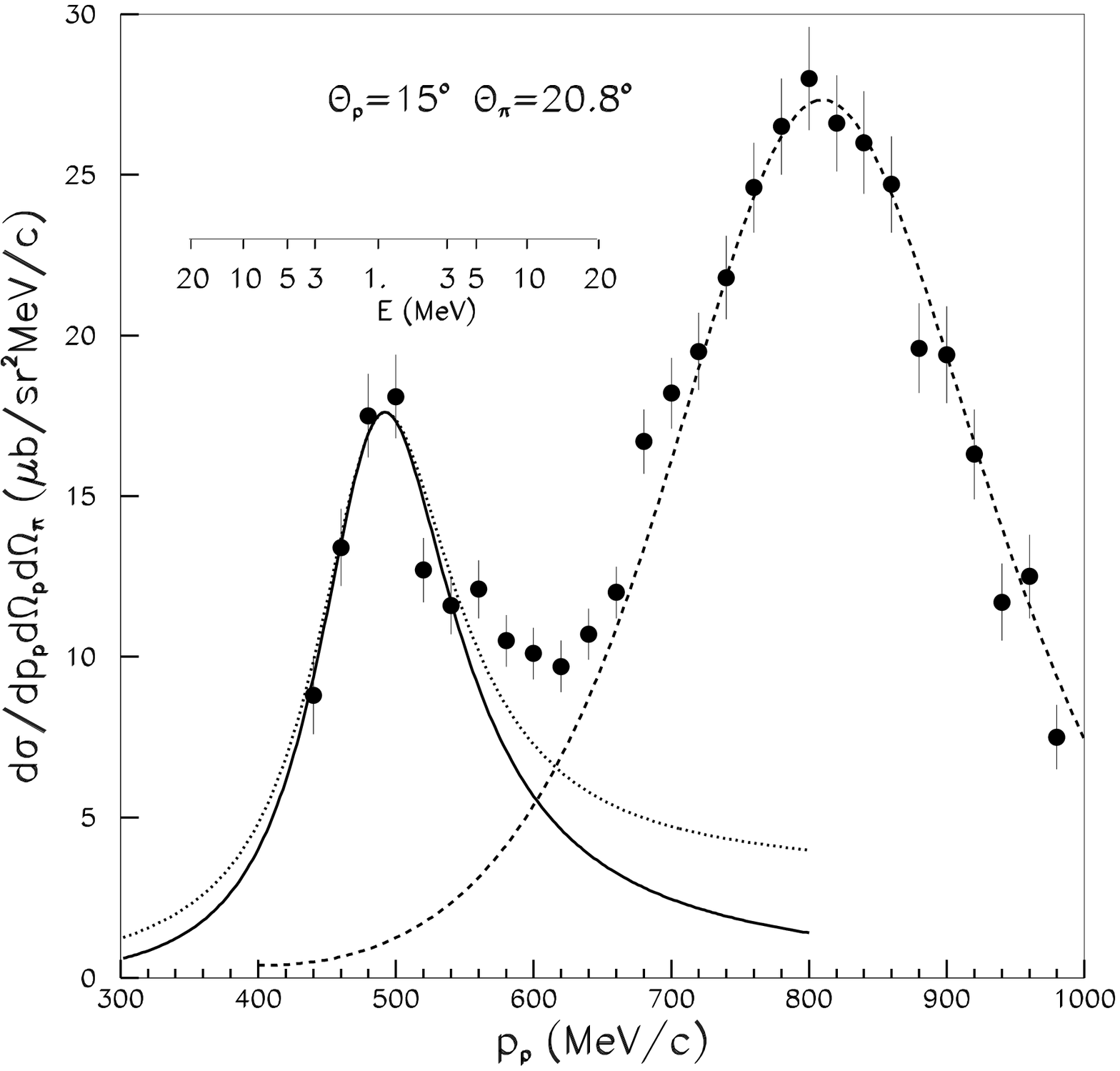}}
\caption{Differential cross section for the $pp\to pn\,\pi^+$
reaction at 800~MeV at fixed proton and pion laboratory angles of
($\theta_p,\theta_{\pi})=(15^{\circ},20.8^{\circ}$) as a function
of the measured proton momentum~\protect\cite{HG}.
The $pp\to \Delta N$ peak on the right can be described
in a one-meson-exchange model (dashed curve)~\protect\cite{Yuri}.
The peak on the left is a reflection of the strong $np$ fsi,
which can be calculated (solid curve) from the $pp\to d\,\pi^+$
cross section, as discussed in the text. A scale is given showing the 
$np$ excitation energy at the geometry corresponding to the centres 
of the counters. For comparison, the triplet fsi factor in the
Goldberger and Watson form~\protect\cite{GW}, multiplied by the
phase space factor of eq.~(\protect\ref{PS}) and normalised to the 
peak, is shown as the dotted curve.}
\end{center}
\label{fig1}
\end{figure}
%%%%%%%%%%%%%%%%%%%%%%%%%%%%%%%%%%%%%%%%%%%%%%%%%%%%%%%%%%%%%%%%%%%%%%
%%%%%%%%%%%%%%%%%%%%%%%%%%%%%%%%%%%%%%%%%%%%%%%%%%%%%%%%%%%%%%%%%%%%%%
\begin{figure}[t]
\begin{center}
\vspace*{-4cm}
\mbox{\epsfxsize=5.5in \epsfbox{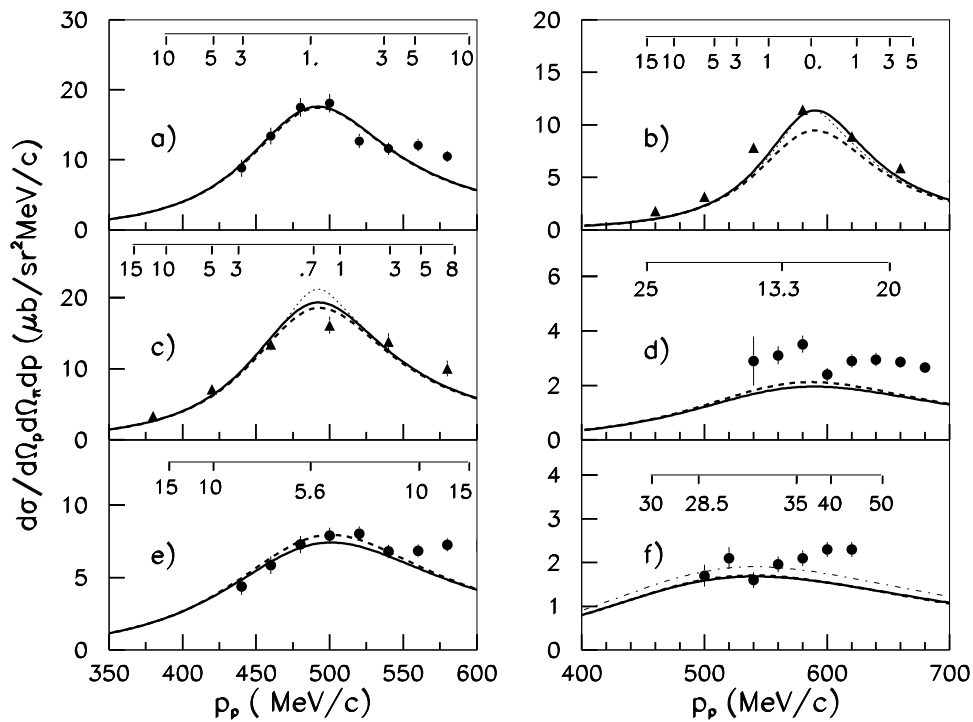}}
\caption{Differential cross section for the $pp\to pn\,\pi^+$
reaction at 800~MeV in the neighbourhood of the $np$ fsi region for
different angular positions, 
a)~$(\theta_p,\theta_{\pi})=(15^{\circ},20.8^{\circ})$, 
b)~$(14.5^{\circ},42^{\circ})$,
c)~$(14.5^{\circ},21^{\circ})$,
d)~$(25^{\circ},40^{\circ})$,
e)~$(20^{\circ},22^{\circ})$,
f)~$(30^{\circ},28^{\circ})$.
The $np$ excitation energies at the counter centres are indicated;
the number with the decimal corresponds to the minimum value of
$E_{np}$. The experimental data are taken from Refs.~\protect\cite{HG} 
(circles) and \protect\cite{Hancock} (triangles) and it should be
noted that these may differ in normalisation by 15\%. The data
are compared to the spin-triplet final-state predictions of
eq.~(\protect\ref{yuri}) without acceptance corrections (solid curve)
and with (broken curve). An attempt to resolve the apparent
discrepancy in b), through the introduction of a singlet contribution
by eq.~(\protect\ref{singlet}) with $\zeta=0.03$ makes the situation
worse in c) (dotted curves). The dot-dash curve in f) was
obtained using the deuteron mass rather than the invariant mass of
the $np$ system when evaluating the pion production angle
$\theta_{\pi}^*$ in eq.~(\ref{theta}).}
\end{center}
\label{fig2}
\end{figure}
%%%%%%%%%%%%%%%%%%%%%%%%%%%%%%%%%%%%%%%%%%%%%%%%%%%%%%%%%%%%%%%%%%%%%%
%%%%%%%%%%%%%%%%%%%%%%%%%%%%%%%%%%%%%%%%%%%%%%%%%%%%%%%%%%%%%%%%%%%%%%
%\input epsf
\begin{figure}[ht]
\begin{center}
\mbox{\epsfxsize=4in \epsfbox{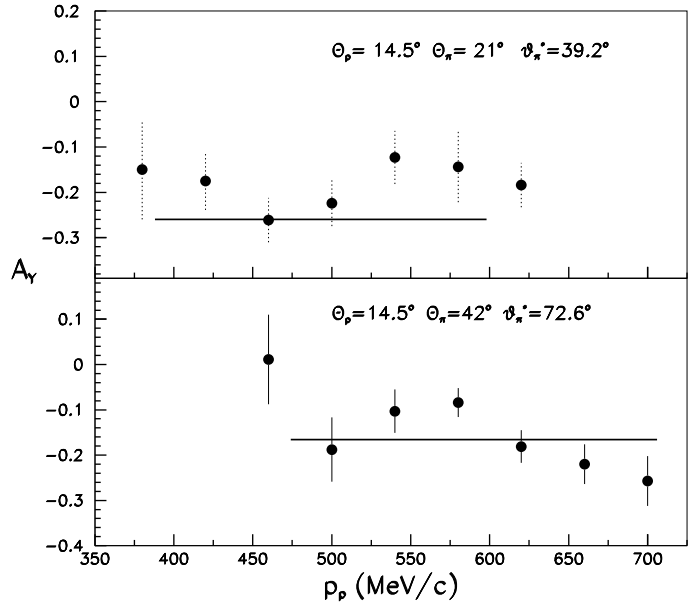}}
\caption{Proton analysing power for the $\vec{p}p\to pn\,\pi^+$
reaction at 800~MeV in the neighbourhood of the $np$ fsi region.
The experimental data of Refs.~\protect\cite{Hancock} are compared
to the SAID predictions of $A_y$ (horizontal lines) for the
$\vec{p}p\to d\,\pi^+$ reaction~\protect\cite{SAID}. It should be noted
that SAID uses the opposite convention for the sign of $A_y$.}
\end{center}
\label{fig3}
\end{figure}
%%%%%%%%%%%%%%%%%%%%%%%%%%%%%%%%%%%%%%%%%%%%%%%%%%%%%%%%%%%%%%%%%%%%%%
%%%%%%%%%%%%%%%%%%%%%%%%%%%%%%%%%%%%%%%%%%%%%%%%%%%%%%%%%%%%%%%%%%%%%%
\end{document}